\documentclass[aps,pra,twocolumn]{revtex4-2}
\usepackage{graphicx}
\usepackage[colorlinks]{hyperref}

\begin{document}

\title{
Superradiant transfer of quantized orbital angular momentum between light and atoms in a ring trap
}

\author{N. Piovella}
\affiliation{Dipartimento di Fisica "Aldo Pontremoli", Universit\`{a} degli Studi di Milano, Via Celoria 16, I-20133 Milano, Italy}
\author{G.R.M. Robb}
\affiliation{SUPA and Department of Physics, University of Strathclyde,Glasgow G4 0NG, Scotland, UK}
\author{R. Bachelard}
\affiliation{Universit\'e C\^ote d'Azur, CNRS, Institut de Physique de Nice, 06560 Valbonne, France}
\affiliation{Instituto de F\'{i}sica de S\~ao Carlos, Universidade de S\~ao Paulo, 13560-970 S\~ao Carlos, SP, Brazil}

\date{\today}
\begin{abstract}
The orbital angular momentum  (OAM) from a laser beam can be coherently transferred  to a Bose-Einstein condensate in a ring trap, in quantized units of $\hbar$.
The light-matter coupling allows for the superradiant transfer of the atoms between the discrete OAM states. Tuning the ring parameters and winding number of the pump light, specific angular momentum states can be populated. This in turn allows control of the emission to generate light with OAM different from that of the pump, as the atomic ring imprints its contribution on the scattered light.
\end{abstract}
\maketitle

{\em Introduction --} Photons possess both translational and spin-angular momentum, where the latter is associated with different polarization
states of the light wave \cite{Beth1935}. Light may also be prepared in states with well-defined orbital angular momentum (OAM), such as in Laguerre-Gaussian (LG) beams~\cite{Allen1992}. Coherent control of the interaction between light and matter facilitates the macroscopic transfer of those quantities from one medium to another, opening up exciting and novel possibilities for quantum information 
engineering, such as storing and transferring light in ensembles of atoms~\cite{krenn2017orbital}.

The spatial modes of the matter field possessing orbital angular momentum are associated with a discrete infinite-dimensional (because of the quantization of the OAM) Hilbert space. Transferring orbital angular momentum from light to a Bose-Einstein condensate  provides a route to entanglement that involves many orthogonal quantum states, rather than just two~\cite{andersen2006quantized,wright2008optical}. Multi-dimensional entangled states could be of considerable importance in the field of quantum information, enabling, for example, a more efficient use of communication channels in quantum cryptography~\cite{fickler2016quantum}.

Recently, it has been shown that cold atoms trapped in a ring trap and driven by a laser beam carrying OAM are set in rotation and acquire bunching in phase~
\cite{gisbert2022superradiant}. Depending on the pump winding number $\ell$ and on the ring radius, different harmonics of the phase become bunched, similarly to the longitudinal configuration of light-matter instabilities~\cite{bonifacio1994}. However, a classical treatment of the azimuthal motion of the atoms means that their OAM is not quantized.

We here investigate the collective coupling between the ring of atoms and light within a quantized approach, so the OAM of the trapped condensate evolves through states with integer numbers of $\hbar$. The light-mediated dipole-dipole interactions between the atoms results in a ``superradiant'' coupling with the light, in analogy with the coupling achieved in cigar-shaped matter waves and linear-momenta states~\cite{Inouye1999}, yet here promoting the transfer of the atoms toward different OAM states. The nonlinear nature of this light-atom dynamics leads to the population of higher OAM states for the atoms and, consequently, for the scattered photons. This population of higher OAM states stems from the presence of higher harmonics in the potential experienced by the atoms. We show that this in turn enables the creation of cat states and the emission of light with OAM different from the pump.

\begin{figure}
\centerline{\includegraphics[width=0.24\textwidth]{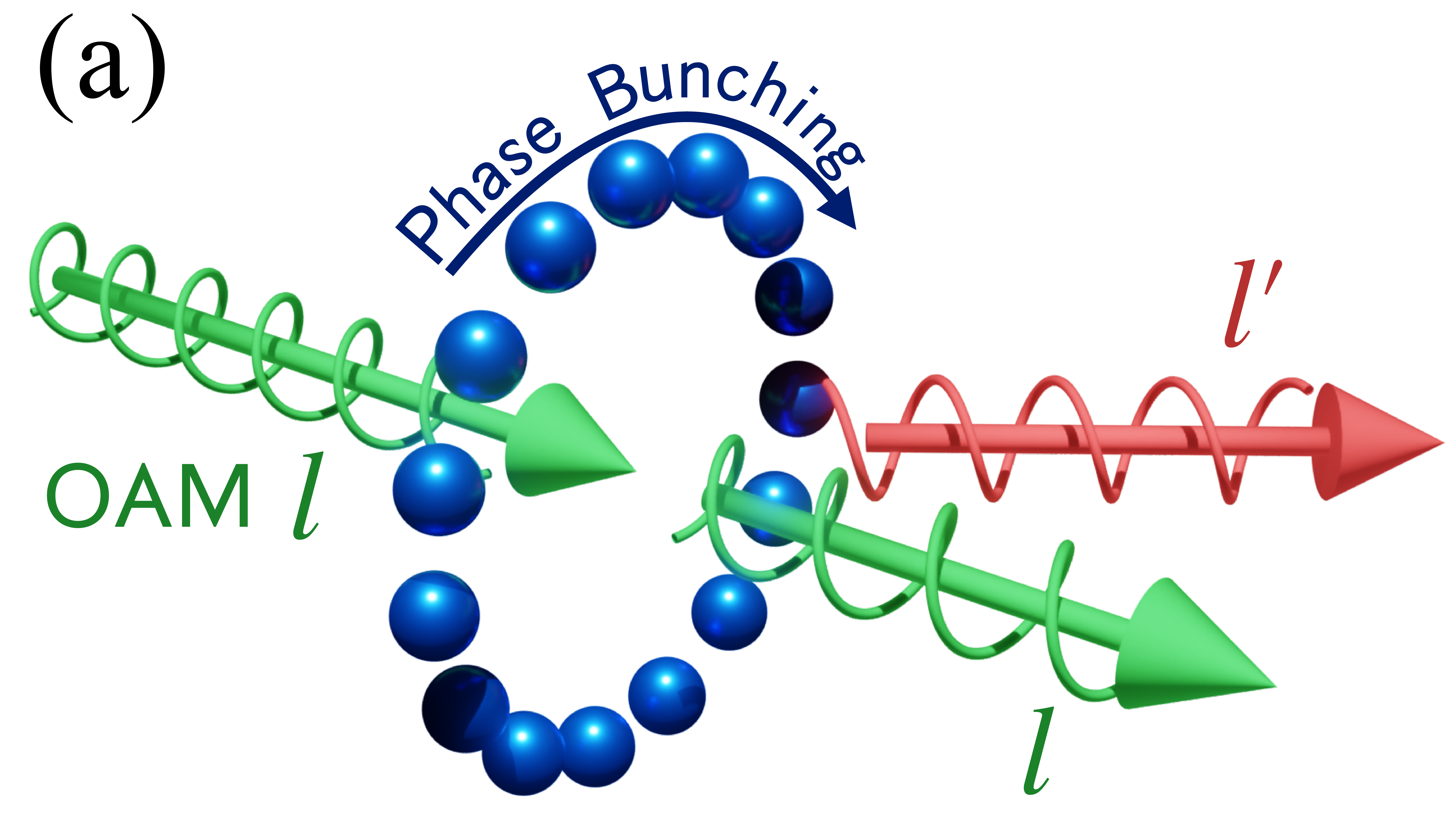}
\includegraphics[width=0.21\textwidth]{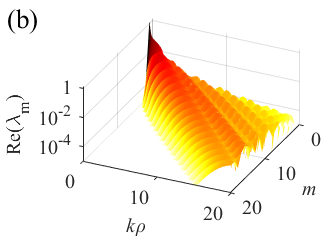}}
\caption{(a) Scheme of the atomic ring pumped with OAM light, so the atoms acquire phase bunching. (b) Growth rate $|\mathrm{Re}(\lambda_m)|$ for the modes $m$, as a function of the ring radius $k_0\rho$. The mode with the strongest rate corresponds to $m\approx k\rho$. Simulations realized with $\gamma=0.2$ and $\epsilon=0.1$.}\label{fig_1}
 \end{figure}

{\em Modelling the light-atom interaction --} 
Let us consider that the atoms are driven resonantly by an OAM laser beam in the mode $\mathrm{LG}_{0\ell}$, with electric field
\begin{equation}
E_0(\rho,\phi,z)=A_0 R(\rho)e^{i(k_0z+\ell\phi-\omega_0t)}+\mathrm{c.c.},\label{OAM:pump}
\end{equation}
with $R(\rho)=(\sqrt{2}\rho/w)^{\ell}\exp(-\rho^2/w^2)$ using cylindrical coordinates, $w$ the beam waist and $\ell$ the pump winding number.
The atoms rotate in a ring trap in the plane perpendicular to the pump propagation axis $z$ (see Fig.~\ref{fig_1}a), and they are described by the matter wave function $\Psi(\phi,\tau)$, where $\phi$ refers to the azimuthal angle. Their dynamics is given by~
\cite{gisbert2022superradiant,SM}
\begin{eqnarray}
i\frac{\partial\Psi(\phi,\tau)}{\partial\tau}&=&-\frac{\partial^2\Psi(\phi,\tau)}{\partial\phi^2}\nonumber\\
&+&\frac{\gamma}{2}\int_0^{2\pi}d\phi'V(\phi-\phi')
|\Psi(\phi',\tau)|^2\Psi(\phi,\tau),\label{eqS}
\end{eqnarray}
with $\tau=\omega_\phi t$ the dimensionless time, $\omega_\phi=\hbar/2M\rho^2$ the angular recoil frequency, $M$ the atom mass and $\rho$ the ring radius. The light-atom coupling is characterized by the parameter $\gamma=N(\Gamma/\omega_\phi)(\Omega_0/2\Delta_0)^2R^2(\rho)$, where $N$ is the atom number, $\Omega_0=dA_0/\hbar$ the pump Rabi frequency, $\Gamma$ the atomic decay rate for the two-level transition at frequency $\omega_a$, $d$ the associated dipole moment, and $\Delta_0=\omega_a-\omega_0$ the pump-atom detuning. The potential $V$ reads
\begin{equation}
V(\varphi)=-\frac{\cos[2k_0\rho q(\varphi)-\ell \varphi]}{k_0\rho q(\varphi)},\label{potential}
\end{equation}
where $q(\varphi)=\sqrt{\sin^2(\varphi/2)+\epsilon^2}$ and $\epsilon$ is a cut-off parameter. The potential $V$ has a finite range due to the multimode emission in vacuum~\citep{Ayllon2019}, and it decreases as $1/r_{jm}$ with the distance $r_{jm}$ between the pair of atoms $(j,m)$. Since the atoms are set on a ring of radius $\rho$, we have $r_{jm}=2\rho|\sin[(\phi_j-\phi_m)/2]|$. The phenomenological cut-off $\epsilon$ in $q(\phi_j-\phi_m)$ has been added to avoid the singularity in the denominator and can be chosen arbitrarily small. 

The wave function $\Psi$ describes the atoms in a BEC state, with the normalization condition $\int_0^{2\pi}|\Psi(\phi,\tau)|^2d\phi=1$. This wave-function can be expanded on the basis of the angular momentum states $|m\rangle$ as $\Psi(\phi,\tau)=(1/\sqrt{2\pi})\sum_m c_m(\tau)\exp(im\phi)$, while the potential is expanded as a Fourier series: $V(\varphi)=\sum_k V_k\exp(ik\varphi)$, with $V_k=(1/2\pi)\int_0^{2\pi}V(\varphi)\exp(-ik\varphi)d\varphi$. By projecting Eq.~(\ref{eqS}) on the state $|m\rangle$, we obtain the dynamical equations for the probability amplitudes $c_m(\tau)$:
\begin{eqnarray}
\frac{dc_m}{d\tau}&=&-im^2c_m-i\frac{\gamma}{2}\sum_k V_k c_{m-k}\sum_n c^*_{n-k}c_n, \label{cm}
\end{eqnarray} 
where the normalization condition now reads $\sum_m |c_m(\tau)|^2=1$.
The radiated intensity emitted by the atoms is
$I(\theta,\phi)=I_1R^2(\rho)N^2|M(\theta,\phi)|^2$, where \cite{SM}
\begin{eqnarray}
M(\theta,\phi)&=&\int_0^{2\pi}e^{-ik_0\rho\sin\theta\cos(\phi-\phi')+i\ell\phi'}|\Psi(\phi')|^2\frac{d\phi'}{2\pi}
\label{M}
\end{eqnarray}
is the dimensionless electric field, $\theta$ is the polar angle, $I_1=(\hbar\omega_0\Gamma/8\pi r^2)(\Omega_0/2\Delta_0)^2$ and $r$ is the distance from the detector. Introducing the $m$-th azimuthal bunching coefficient
\begin{equation}
\Phi_m(\tau)=\sum_{n=-\infty}^{+\infty} c_{n-m}^*(\tau)c_n(\tau),
\end{equation}
the electric field rewrites as the sum of different OAM components, composed by photons with angular momentum $\hbar\ell'=\hbar(\ell+m)$:
\begin{equation}
M(\theta,\phi)=\sum_m(-i)^{\ell+m}J_{\ell+m}(k_0\rho\sin\theta)\Phi_m(\tau) e^{i(\ell+m)\phi},\label{M2}
\end{equation}
where $J_n(x)$ is the Bessel function of the first kind and of order $n$.
The rescaled intensity averaged over the azimuthal angle is $\bar I(\theta)=\sum_m J_{\ell+m}^2(k_0\rho\sin\theta)|\Phi_m|^2$.

{\em Linear stability analysis --} 
Let us now discuss the atom dynamics. We observe that the cubic nonlinearity in Eq.~(\ref{cm}) couples state $m$ with the states $m\pm k$ provided that a $k$-th Fourier component $V_k$ of the potential and a $k$-th azimuthal bunching $\Phi_k$ are present. More quantitatively, the dynamical growth of the azimuthal bunching at harmonic $m$ is obtained by a linear stability analysis of Eq.~(\ref{cm}) around the equilibrium point $\Phi_m^{(0)}=\delta_{m,0}$, which describes uniformly distributed phases. This leads to the complex eigenvalues \cite{SM}
\begin{equation}
\lambda_m=\pm i m\sqrt{m^2+\gamma V_m},\label{lambda}
\end{equation}
where the $m$-th azimuthal bunching $\Phi_m$ grows exponentially as $\exp[\mathrm{Re}(\lambda_m)\tau]$. Hence, the exponential growth of the azimuthal bunching $\Phi_m$ requires a finite imaginary part of the $m$-th Fourier component of the potential, $V_m$. For $\ell=0$ the potential $V(\varphi)$ is an even function of $\varphi$ and $V_m$ is real, hence the system is unstable only in presence of OAM light, with $\ell\neq 0$. Intuitively, the OAM pump exerts on the atoms a torque forcing them to rotate. Since the atomic angular momentum is quantized, the atoms rotate with a discrete angular momentum, whose value is determined by the anharmonicity of the potential $V$. 

When uniformly distributed in phase, all the atoms rotate with the same angular velocity and no macroscopic rotation is visible. However, in presence of an OAM beam, the combined effect of the pump and scattered beam creates a self-consistent potential which modulates the azimuthal atomic phases. As a result, atoms are coupled by the scattered photons, recoiling in phase as angular momentum is transferred from the light. Because the beam-matter coupling grows with the bunching in phase, an instability develops and both the bunching in phase and the scattered intensity grow exponentially until the moment when the atoms are transferred to a new OAM state $|m\rangle$. In the case of a subwavelength trap, the ring of atoms essentially behaves as a point-like dipole and the matter is transferred to the state $|\ell\rangle$, where $\hbar\ell$ is the angular momentum of the pump photons. However, increasing the ring radius $\rho$, the potential presents higher harmonics and is able to transfer the atoms to OAM states with $m$ larger than $\ell$.

\begin{figure*}
	\centering 
	\includegraphics[width=\textwidth]{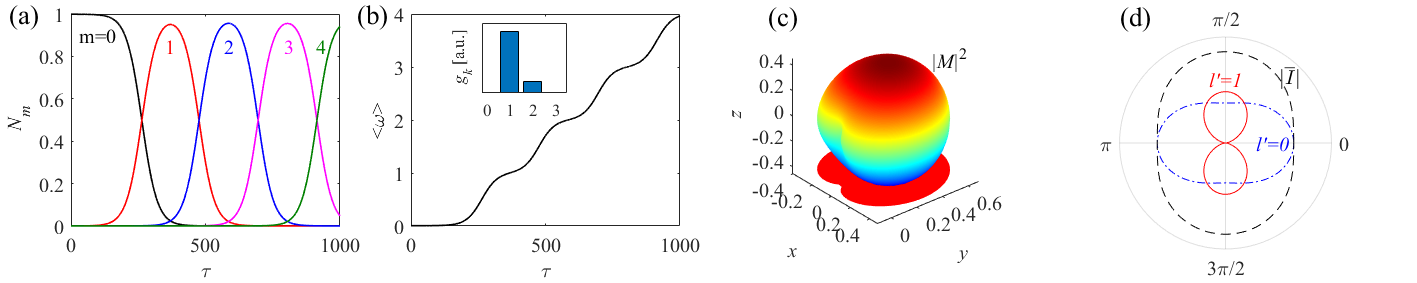}
	\caption{(a) Populations $N_m$ and (b) average angular velocity $\langle\omega\rangle$, in units of $\omega_\phi$, as a function of the dimensionless time $\tau$. (c) Dimensionless scattered intensity $|M(\theta,\phi)|^2$ and (d) $\bar I(\theta)$ vs $\theta$ for $k_0\rho=1$, $\ell=1$ and $\gamma=0.05$; The blue line is the component with $\ell'=0$, the red line is the component with  $\ell'=1$ and the dashed black line is the total intensity. for $\gamma=0.05$, $k_0\rho=1$ and $\ell=1$. The inset in (b) presents the rate coefficients $g_k$ (in a.u.) as a function of $k$, showing that the dominant transition is for $k=1$.}
	\label{fig2}
\end{figure*}

According to Eq.~(\ref{lambda}), the system admits two different regimes: a \textit{classical regime} when $\gamma |V_m|\gg m^2$ and thus $\lambda_m\approx\pm i m\sqrt{\gamma V_m}$, which is characterized by an exponential growth rate proportional to $\sqrt{N}$, and a \textit{quantum regime}, when $\gamma |V_m|\ll m^2$ so that $\lambda_m\approx\pm i(m^2+\gamma V_m/2)$, which presents a rate proportional to $N$~\cite{robb2005semiclassical}. In the classical regime, both the states $|m\rangle$ and $|-m\rangle$ become populated almost simultaneously. Differently, in the quantum regime, the dynamical term $\exp(\pm im^2\tau)$ corresponds to a phase shift proportional to the rotational energy $L_z^2/2M\rho^2=\hbar \omega_\phi m^2$, with orbital angular momentum $L_z=\hbar m$. This energy shift, which stems from energy and angular momentum conservation in the photon-atom scattering process, hinders the occupation of one of the two states $|\pm m\rangle$ (which one is populated depends on the sign of $\mathrm{Im}(V_m)$). This leads to the scaling of the growth rate as $N$ for the quantum regime~\cite{robb2005semiclassical}.

The scattering of the pump light into different OAM states can be deduced from the eigenspectrum, whose real part $|\mathrm{Re}(\lambda_m)|$ is presented in Fig.~\ref{fig_1}b for the quantum regime ($\gamma=0.2$), as a function of the radius $k_0\rho$ and of $m$. The fact that for increasing $k_0\rho$ the maximum occurs at values of $m$ which scales linearly with $k_0\rho$ implies that larger ring are able to populate higher OAM states of the condensate. 
This is a notable difference with the case of the one-dimensional quantum collective atomic recoil lasing (CARL)~\cite{bonifacio1994, Piovella2001b, Slama2007}, where only transitions between adjacent linear momentum states are possible. This is due to the fact that CARL is characterized by a sinusoidal potential at the light wavelength, which results from the interference between the pump and the backward scattered modes. In the ring trap, the presence of higher harmonics in potential~(\ref{potential}) allows the atoms to bunch at higher OAM states but also, as we shall see later, to scatter photons of different OAM in different directions.

{\em Rate equations --} 
In the quantum regime, when $\gamma |V_m|\ll m^2$, the atoms are transferred to OAM states $|m\rangle$ through a superradiant cascade which involves, essentially, only two states at the time. Then, Eqs.~(\ref{cm}) are approximated  by the following rate equations for the state populations $N_m=|c_m|^2$:
\begin{equation}
\frac{dN_m}{d\tau}=\left[ \sum_{k=1}^{m}g_k N_{m-k}
-\sum_{k=1}^{+\infty} g_k N_{m+k}\right]N_m,\label{rate}
\end{equation}
where $g_k=\gamma |\mathrm{Im} (V_k)|$ are the rate coefficients, and with the normalization condition $\sum_m N_m=1$. These coupled equations generalize the superradiant transition between two states $N_0$ and $N_k$ when only a single coefficient $g_k$ is present, and it solves as $N_{0,k}(\tau)=(1/2)\{1\mp\tanh[g_k(\tau-\tau_0)/2]\}$, with $\tau_0=(1/g_k)\ln[2/\sqrt{N_k(0)}]$ the delay time.
In the superradiant cascade the coherence of the matter wave function is preserved and the phases of the complex amplitudes $c_m=\sqrt{N_m}e^{i\phi_m}$ are determined by the equations
 \begin{equation}
\frac{d\phi_m}{d\tau}=-(m^2+\gamma V_0)-\left[ \sum_{k=1}^{m}\alpha_k N_{m-k}
+\sum_{k=1}^{+\infty}\alpha_k N_{m+k}\right],
\end{equation}
with $\alpha_k=(\gamma/2)\mathrm{Re} (V_k)$.

For a small ring radius, $k_0\rho\ll 1$, the atoms are transferred sequentially through the OAM states by steps of $\Delta m=\ell$. This is illustrated in Fig.~\ref{fig2}, which depicts the temporal evolution of the populations $N_m$ for the first OAM states (see panel (a)) and the average angular velocity $\langle\omega\rangle=\sum_m m N_m$ (see panel (b)). 
The angular distribution of the scattered field intensity $|M(\theta,\phi)|^2$ is plotted in Fig.~\ref{fig2}(c) at the time when the bunching is maximum ($|\Phi_1|\approx 1/2$), while panel (d) shows the total average intensity $\bar I(\theta)$ as a function of the polar angle $\theta$ (dashed line), together with the main OAM components. In this case the field (\ref{M}) is approximated by
$M(\theta,\phi)\approx -iJ_1(k_0\rho\sin\theta)\exp(i\phi)+\Phi_1^*J_0(k_0\rho\sin\theta)$, with $|\Phi_1|\approx 1/2$ and where we have neglected the term proportional to $J_2(\sin\theta)$. 
 The average intensity is $\bar I(\theta)\approx J_1^2(k_0\rho\sin\theta)+|\Phi_1|^2J_0^2(k_0\rho\sin\theta)$.
 In Fig. \ref{fig2} we can observe the component $\ell'=1$ (red line), the component $\ell'=0$ (blue line) and the total intensity (dashed black line). The light with the same orbital angular momentum of the pump $\ell=1$ is emitted in the perpendicular direction, whereas the light with $\ell=0$ is emitted forward and backward within a large diffraction angle. The total intensity is almost isotropic, as expected for a small ring.

\begin{figure}
\centerline{\includegraphics[width=0.5\textwidth]{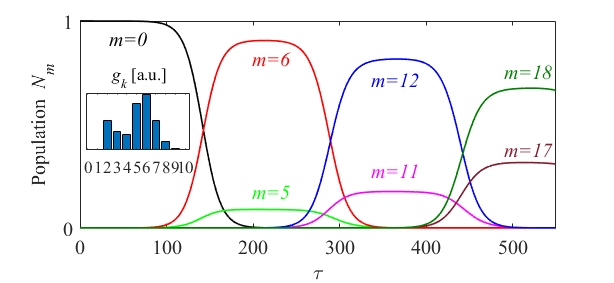}}
\caption{Populations $N_m$ as a function of the dimensionless time $\tau$. Simulations realized for $k_0\rho=5.605$, $\ell=2$ and $\gamma=1$. The inset presents the rate coefficients $g_k$ (in a.u.) as a function of  $k$, showing that the dominant transitions are for $k=5$ and $k=6$.}\label{fig_3}
\end{figure}

{\em Populating higher OAM states --} Atomic rings larger than the wavelength are able to populate directly atomic states with large OAM, taking advantage of the higher harmonics of the potential $V$. In Fig.~\ref{fig_3} we illustrate how by choosing parameters such that the potential favors the coupling to several higher OAM states, the system can be cast in a macroscopic superposition. The pump with $\ell=2$ transfers most of the atomic population from $m=0$ to $m=6$, and to $m=5$ to a lesser extent, due to the set of coefficients $g_k$ among which $g_6$ and $g_5$ are highest (the parameters have been chosen so $g_1=0$, so transitions with  $\Delta m=1$ are suppressed). The first transition sees a transfer of population from $m=0$ to $m=6$ ($N_6=91\%$) and to $m=5$ ($N_5=9\%$). During the next transitions, the atoms are transferred to $m=12$ and $m=11$, then to $m=18$ and $m=17$, and so on, since the specific set of parameters favors transitions with $\Delta m=6$. Thus, after two transition channels are open (that is, with $\Delta m=5$ and $6$), a macroscopic superposition of two rotational states, is created. Throughout the successive transitions toward higher phase bunching states, superpositions of two states persist. The possibility that these superposition states are Schrodinger cat states will be investigated elsewhere.
 
\begin{figure*}
\centering 
\includegraphics[width=\textwidth]{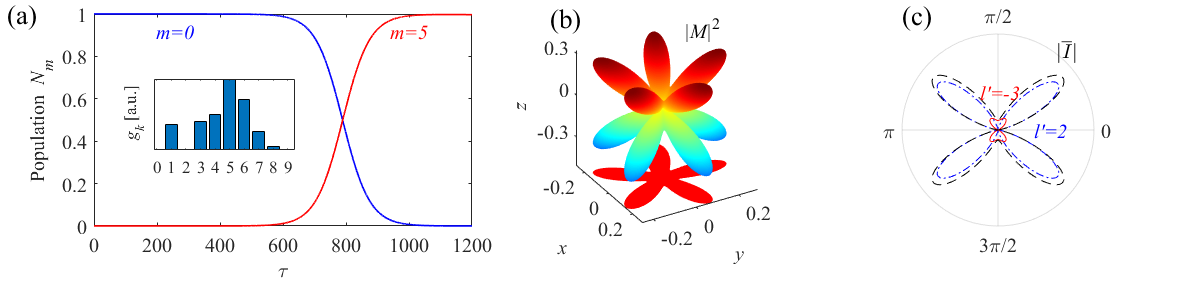}
\caption{(a) Population of the state $m=0$  and $m=5$ as a function of the dimensionless time $\tau$. The inset shows the coefficients $g_k$. (b) Angular distribution of the radiation intensity $|M(\theta,\phi)|^2$, at time $\tau=800$. (c) Dimensionless intensity $\bar I(\theta)$ averaged over the azimuthal angle $\phi$, as a function of the polar angle $\theta$, at time $\tau=800$. The dashed black line stands for the total intensity, the blue line for the component $\ell'=2$ and the red line for the component $\ell'=-3$: the latter is the only component present at $\theta=\pi/2$. Simulations realized for $k_0\rho=5$, $\ell=2$ and $\gamma=0.2$.}
	\label{fig4}
\end{figure*}

{\em Tailoring the light emission --} The dynamical evolution to higher OAM atomic states in turn results in the emission of light whose angular momentum differs from the pump one. This can be a tool either for reading the atomic state, or for producing light with a specific OAM. In particular, one may choose the ring radius such that no photons with the same OAM of the pump are emitted in a particular direction, so light with a different OAM can be collected more efficiently. As an example, we show in Fig.~\ref{fig4}(a) how an atomic ring with radius $k_0\rho=5$, along with a pump of winding number $\ell=2$, is transferred to the OAM state $m=5$.

This bunching in phase leads to five-lobe radiation pattern in the azimuthal direction, see panel (b).
Because the Bessel function $J_2(k_0\rho\sin\theta)$ vanishes for $k_0\rho=5$ and $\theta=\pi/2$, no photons with $\ell'=\ell=2$ are emitted in the transverse direction. At this angle, only the component with $\ell'=l+m=-3$ of the scattered light is present. Thus, in this specific direction, light with a well-defined OAM, which is different from that of the pump, can be collected.

In conclusion, we have demonstrated how orbital angular momentum can be transferred coherently from a single Laguerre-Gaussian to a Bose-Einstein condensate in a ring trap. The spontaneous formation of the azimuthal bunching of ultracold atoms in the ring trap is at the core of the process. The process is similar to the quantum regime of collective atomic recoil lasing~\cite{bonifacio1994, Piovella2001b, Slama2007} and to the superradiant Rayleigh scattering~\cite{Inouye1999}, but with transfer of orbital angular momentum in units of $\hbar$ instead of linear momentum in units of $\hbar k$. Differently from these linear configurations, the many harmonics of the potential here allow to promote the atoms to one or more higher OAM states. This mechanism is particularly promising for the creation of states of matter with tailored OAM~\cite{robb2012superradiant,Das_2016,kumar2021cavity}: in the present scheme, the choice of the ring radius, by tuning the trap frequencies, allows one to manipulate the relative populations of different angular momentum states. The emission of photons with angular momentum different from that of the pump and with a specific spatial pattern offers the possibility to either use this light to read the atomic state, or to use the atomic ring to produce light with a new OAM. The atomic ring may also be used as a beam splitter, dividing the light into two or more topological states, so robust phase qubits can be generated. In this context, one may take advantage of the complete orthonormal basis formed by the OAM states to design more elaborate quantum states for quantum information processing~\cite{kapale2005vortex}. 

The proposed setup correspond to ring traps with radii of a few optical wavelengths, or subwavelength for the case $\rho<\lambda$ (yet this latter case does not produce efficiently higher OAM light).
Recent advances in the control of quantum gases have seen the development of atom ring traps formed from both magnetic and optical dipole potential, and more recent implementations using hybrids of both, down to $\rho=9\mu m$ ring traps~\cite{amico2021roadmap,eckel2014hysteresis}. Although much more challenging, nanometer-scale traps are also currently under investigation~\cite{richberg2021optical}.

\section*{Acknowledgments} 
We thank Andr\'e Cidrim for his help with editing figures. R.B. has received support from the S\~ao Paulo Research Foundation (FAPESP) through Grants No.2019/12842-2, 2019/13143-0 and 2018/15554-5, as well as from the National Council for Scientific and Technological Development (CNPq) Grant Nos. 409946/2018-4 and 313886/2020-2. Part of this work was performed in the framework of the European Training Network ColOpt, which is funded by the European Union (EU) Horizon 2020 program under the Marie Sklodowska-Curie action, grant agreement No. 721465.

\bibliography{Bibliography}

\end{document}


\title{Supplementary Material to "Superradiant transfer of quantized orbital angular momentum between light and atoms in a ring trap"}

\author{N. Piovella}
\affiliation{Dipartimento di Fisica "Aldo Pontremoli", Universit\`{a} degli Studi di Milano, Via Celoria 16, I-20133 Milano, Italy}
\author{G.R.M. Robb}
\affiliation{SUPA and Department of Physics, University of Strathclyde,Glasgow G4 0NG, Scotland, UK}
\author{R. Bachelard}
\affiliation{Universit\'e C\^ote d'Azur, CNRS, Institut de Physique de Nice, 06560 Valbonne, France}
\affiliation{Instituto de F\'{i}sica de S\~ao Carlos, Universidade de S\~ao Paulo, 13560-970 S\~ao Carlos, SP, Brazil}

\date{\today}
\maketitle

\section{Model}

We here derive the Hamiltonian describing the dynamics of $N$ atoms in a ring trap of radius $\rho$, in a plane perpendicular to the propagation direction of the incident OAM laser beam. The photons are scattered by the atoms in vacuum free space. The first step consists in  adiabatically eliminating the internal degrees of freedom, assuming a weak pump and a large atom-radiation frequency detuning (as compared to the transition linewidth). Furthermore, in free space the scattered field 
is also adiabatically eliminated, which eventually leads to a set of coupled equations for the atomic azimuthal phases. 

Let us assume a pump consisting of a hollow Laguerre-Gaussian mode $LG_{0\ell}$ propagating along the $z$-axis and linearly polarized. Its electric field component, expressed in cylindrical coordinates, is~\cite{yao2011orbital}:
\begin{equation}
E_0(\rho,\phi,z)=A_0 R(\rho)e^{i(k_0z+\ell\phi-\omega_0t)}+\mathrm{c.c.},\label{OAM:pump}
\end{equation}
where c.c. stands for the complex conjugate. Without loss of generality, we assume that $\ell>0$, so that 
\begin{eqnarray}
R(\rho)&=&e^{-\rho^2/w^2}\left(\frac{\sqrt{2}\rho}{w}\right)^{\ell},\label{R(r)}
\end{eqnarray}
where $w$ is the beam waist, $A_0=\sqrt{2^{\ell+1}/l!}\sqrt{P_0/(\pi w^2\epsilon_0 c)}$ and $P_0$ is the mode power. 
We assume $N$ two-level atoms with internal states $|g_j\rangle$ and $|e_j\rangle$ (transition frequency $\omega_a$), confined in a ring trap with radius $\rho$ and in the transverse plane $z=0$, so the atomic positions and momenta are $\mathbf{r}_j=\rho(\cos\phi_j,\sin\phi_j,0)$ and $\mathbf{p}_j=M\rho\dot\phi_j(-\sin\phi_j,\cos\phi_j,0)$ (where $\phi_j$ is the azimuthal angle and $\dot\phi_j=d\phi_j/dt$, with $j=1,\dots,N$). The atom mass is $M$, the dipole moment $d$. The atoms interact with the pump OAM field (\ref{OAM:pump}) and scatter photons in the vacuum modes, so the system is described by the following Hamiltonian:
\begin{equation}
H=T+H_L+H_V.\label{H}
\end{equation}
with
\begin{eqnarray}
T=\sum_{j=1}^N\frac{\mathbf{p}_j^2}{2M}
\end{eqnarray}
the kinetic energy. Hamiltonian $H_L$ describes the interaction of the atoms with the pump laser,
\begin{eqnarray}\label{HL}
    H_L =\frac{\hbar\Omega_0}{2}R(\rho)\sum_{j=1}^N \left[\sigma_{j}^- 
    e^{i\Delta_0t-i\ell\phi_j}+\mathrm{h.c.}\right],
\end{eqnarray}
where $\mathrm{h.c.}$ stands for the hermitian conjugate, $\Omega_0=dA_0/\hbar$ is the Rabi frequency and $\Delta_0=\omega_0-\omega_a$ the pump-atom detuning.
The internal dynamics of the two-level atoms is described by the operators
$\sigma_{j}^z=|e_j\rangle \langle e_j|- |g_j\rangle \langle g_j|$,~
$\sigma_{j}^+=|e_j\rangle\langle g_j|$ and
$\sigma_{j}^-=|g_j\rangle\langle e_j|$. Finally, $H_V$ describes the interaction between the atoms and the vacuum-mode field:
\begin{eqnarray}\label{HV}
    H_V =\hbar\sum_{\mathbf{k}}g_k\sum_{j=1}^N\left[a_{\mathbf{k}}^\dagger \sigma_{j}^- e^{i\Delta_kt-i\mathbf{k}\cdot\mathbf{r}_j}
        +\mathrm{h.c.}\right].
\end{eqnarray}
The operators $a_\mathbf{k}$ here describes the annihilation of a photon from vacuum mode $\mathbf{k}$, with wavevector $\mathbf{k}$ and frequency $\omega_k$, coupling rate $g_k=d[\omega_k/(2\hbar\epsilon_0 V_{\mathrm{ph}})]^{1/2}$,
and detuning $\Delta_k=\omega_k-\omega_a$ from the atomic transition. $V_{\mathrm{ph}}$ is the quantization volume. We have here disregarded polarization and short-range effects, adopting a scalar model for the radiation field.

The internal degrees of freedom can be adiabatically eliminated assuming a weak driving field (so the population of the excited state can be neglected, $\sigma_{j}^z\approx -1$) and a large detuning, $\Delta_0\gg\Gamma$, where   $\Gamma=d^2k^3/2\pi\epsilon_0\hbar$ is the spontaneous decay rate of the atomic transition. This results in
\begin{eqnarray}
\sigma_j^- &\approx & \frac{1}{\Delta_0}\left[\frac{\Omega_0}{2}R(\rho_j)e^{ik_0z_j+i\ell\phi_j}
  +\sum_{\mathbf{k}}g_k
 a_{\mathbf{k}}e^{-i(\omega_k-\omega_0)t+i\mathbf{k}\cdot \mathbf{r}_j}\right]e^{-i\Delta_0 t}\label{s1:adia}.
 \end{eqnarray}
 The adiabatic elimination of the internal degrees of freedom leads to the following effective Hamiltonian:
 \begin{eqnarray}\label{H_adia}
    H' =\sum_{j=1}^N\left\{\frac{L_{zj}^2}{2M\rho^2}+\frac{\hbar\Omega_0}{2\Delta_0}R(\rho)\sum_{\mathbf{k}} g_k
    \left[a_{\mathbf{k}}^\dagger e^{i(\omega_k-\omega_0)t-i\mathbf{k}\cdot\mathbf{r}_j+i\ell\phi_j}
        +\mathrm{h.c.}\right]\right\}=\sum_{j=1}^N H_j,
\end{eqnarray}
where we have neglected multiple-scattering processes (second-order terms in $a_{\mathbf{k}}$) and introduced the $z$-component of the angular momentum of the $j$th atom, $L_{zj}=M\rho^2\dot\phi_j$.

The Heisenberg equation for the vacuum field is
\begin{eqnarray}
  \dot a_{\mathbf{k}} &=& -ig_k\frac{\Omega_0}{2\Delta_0}R(\rho)e^{i(\omega_k-\omega_0)t} \sum_{j=1}^N
e^{-i\mathbf{k}\cdot \mathbf{r}_j+i\ell\phi_j}, \label{D:A:j}
 \end{eqnarray}
reminding that in free space, the light is scattered in all vacuum modes. Following Ref.~\cite{Ayllon2019}, we eliminate the scattered field by integrating  Eq.~(\ref{D:A:j}) to obtain
\begin{equation}
a_{\mathbf{k}}(t) = a_{\mathbf{k}}(0)
-ig_k\frac{\Omega_0}{2\Delta_0}R(\rho)e^{i(\omega_k-\omega_0)t}\sum_{j=1}^N\int_0^te^{i\ell\phi_j(t-\tau)-i\mathbf{k}\cdot\mathbf{r}_j(t-\tau)}e^{-i(\omega_k-\omega_0)\tau}d\tau.\label{ak:int}
\end{equation}
We neglect the first source term in Eq.~(\ref{ak:int}), which is associated to vacuum fluctuations. The second term describes Rayleigh scattering. We make the Markov approximation, assuming that the term under the integral can be rewritten $\phi_j(t-\tau)\approx \phi_j(t)$ and 
$\mathbf{r}_j(t-\tau)\approx \mathbf{r}_j(t)$, and taking the $t\rightarrow\infty$ limit for the integral, so that
\begin{equation}
a_{\mathbf{k}}(t) =
-ig_k\frac{\Omega_0}{2\Delta_0}R(\rho)e^{i(\omega_k-\omega_0)t}\sum_{j=1}^Ne^{i\ell\phi_j(t)-i\mathbf{k}\cdot\mathbf{r}_j(t)}\int_0^\infty e^{-i(\omega_k-\omega_0)\tau}d\tau.\label{ak:int:2}
\end{equation}
We then insert Eq.~(\ref{ak:int:2}) into Eq.~(\ref{H_adia}) and we transform the sum over $\mathbf{k}$ into an integral. Then, using the identity
\begin{eqnarray}
\int d\mathbf{k}e^{-i\mathbf{k}\cdot(\mathbf{r}_j-\mathbf{r}_m)}e^{ick\tau}&=&i\frac{4\pi^2 k_0}{cr_{jm}}
\left[\delta(\tau-r_{jm}/c)-\delta(\tau+r_{jm}/c)\right],
\end{eqnarray}
where $r_{jm}=|\mathbf{r}_{j}-\mathbf{r}_{m}|$, we obtain
\begin{eqnarray}\label{H_adia:3}
    H_j =\frac{L_{zj}^2}{2M\rho^2}-\hbar\Gamma\left(\frac{\Omega_0}{2\Delta_0}\right)^2R^2(\rho)\sum_{m=1}^N 
    \frac{\cos(k_0r_{jm}-\ell\phi_{jm})}{k_0r_{jm}}.
\end{eqnarray}
The distance between two atoms trapped in the ring of radius $\rho$ is $r_{jm}=2\rho|\sin[(\phi_j-\phi_m)/2]|$. In order to avoid the singularity in the denominator of the potential for $\phi_j=\phi_m$, we introduce a cut-off $\epsilon$, which can be chosen arbitrarily small, and the function $q(x)=\sqrt{\sin^2(x/2)+\epsilon^2}$. Hence, we obtain
\begin{eqnarray}\label{H_eff}
    H_j =\frac{L_{zj}^2}{2M\rho^2}+\frac{\hbar\Gamma}{2}\left(\frac{\Omega_0}{2\Delta_0}\right)^2R^2(\rho)\sum_{m=1}^N 
    V(\phi_j-\phi_m),
\end{eqnarray}
where
\begin{equation}
V(x)=-\frac{\cos[2k_0\rho q(x)-\ell x]}{k_0\rho q(x)}.
\end{equation}
We observe that since the two-body interaction potential potential $V(\phi_j-\phi_m)$ is not symmetric, the dynamics of the $N$ atoms in the ring  cannot be derived from a Hamiltonian \cite{gisbert2022superradiant}. Nevertheless, $H_j$ can be considered an effective Hamiltonian for the $j$th atom~\cite{gisbert2022superradiant}, to which are associated the following dynamical equations:
\begin{eqnarray}
\frac{d\phi_j}{dt} &=& \frac{\partial H_j}{\partial L_{zj}}=\frac{L_{zj}}{M\rho^2}\\
\frac{dL_{zj}}{dt} &=& -\frac{\partial H_j}{\partial\phi_j}=-
\frac{\hbar\Gamma}{2}\left(\frac{\Omega_0}{2\Delta_0}\right)^2R^2(\rho)\frac{\partial}{\partial\phi_j} \sum_{m=1}^N 
    V(\phi_j-\phi_m).
\end{eqnarray}

From the single-particle Hamiltonian (\ref{H_eff}), we can write the following nonlinear Schr\"{o}dinger equation for the matter wavefunction $\Psi(\phi,t)$, using the substitution $\phi_j\rightarrow \phi$, $L_{zj}\rightarrow -i\hbar(\partial/\partial\phi)$ and $\sum_m\rightarrow N\int_0^{2\pi}d\phi'|\Psi(\phi')|^2$,
\begin{eqnarray}
i\hbar\frac{\partial\Psi(\phi,t)}{\partial t}&=&-\frac{\hbar^2}{2M\rho^2}\frac{\partial^2\Psi(\phi,t)}{\partial\phi^2}+\frac{\hbar\Gamma}{2}\left(\frac{\Omega_0 R(\rho)}{2\Delta_0}\right)^2N\int_0^{2\pi}d\phi'
    V(\phi-\phi')|\Psi(\phi',t))|^2\Psi(\phi,t),
\end{eqnarray}
with the normalization condition
\begin{equation}
\int_0^{2\pi}|\Psi(\phi,t)|^2=1.
\end{equation}
Introducing the angular recoil frequency $\omega_\phi=\hbar/2M\rho^2$ and the dimensionless time $\tau=\omega_\phi t$, we obtain the nonlinear Schr\"{o}dinger equation (2) of the main text,
\begin{eqnarray}
i\frac{\partial\Psi(\phi,\tau)}{\partial\tau}&=&-\frac{\partial^2\Psi(\phi,\tau)}{\partial\phi^2}
+\frac{\gamma}{2}\int_0^{2\pi}d\phi'
    V(\phi-\phi')|\Psi(\phi',\tau)|^2\Psi(\phi,\tau)
\end{eqnarray}
where $\gamma=N(\Gamma/\omega_\phi)(\Omega_0/2\Delta_0)^2R^2(\rho)$. We observe that a rigorous second-quantization treatment of this system would require a master equation approach~\cite{breuer2002theory}, which will be the object of future studies. 

\section{Radiation field -- Derivation of Eq.(5)}

The radiation field amplitude scattered by the atoms is given by
\begin{equation}
 E_s(\mathbf{r},t)=i\frac{V_{\mathrm{ph}}}{8\pi^3}\int d\mathbf{k} {\cal E}_k a_\mathbf{k}(t)e^{i\mathbf{k}\cdot\mathbf{r}-i\omega_k t},
 \label{Es}
 \end{equation}
with ${\cal E}_k=(\hbar\omega_k/2\epsilon_0 V_{\mathrm{ph}})^{1/2}$ the 'single-photon' electric field.
Using Eq.~(\ref{ak:int}), neglecting the fluctuation term, transforming the sum over $\mathbf{k}$ into an integral, and performing the integration over the angular part of $\mathbf{k}$, we obtain
\begin{eqnarray}
 E_s(\mathbf{r},t)
&=&\frac{V_{\mathrm{ph}}}{8\pi^3}\frac{\Omega_0}{2\Delta_0}R(\rho)e^{-i\omega_0 t}\sum_{j=1}^N\int_0^t d\tau 
\frac{e^{i\omega_0\tau+i\ell\phi_j(t-\tau)}}{|\mathbf{r}_j(t-\tau)-\mathbf{r}|}
\int_0^\infty dk kg_k{\cal E}_{k}\sin(k|\mathbf{r}_j(t-\tau)-\mathbf{r}|)e^{-ick\tau}.
 \end{eqnarray} 
The frequency of the scattered field is centered on the incidence laser frequency $\omega_0$. 
integral in $\tau$ is not negligible. 
In the integral, we therefore replace $k$ and $g_k$ by $k_0$ and $g_{k_0}$, respectively, and extend
the lower limit in the $k$-integration to $-\infty$:
\begin{eqnarray}
 E_s(\mathbf{r},t)&=&\frac{V_{\mathrm{ph}}}{2\pi^2}\frac{\Omega_0}{2\Delta_0}R(\rho)
 k_0g_{k_{0}}{\cal E}_{k_0}e^{-i\omega_0 t}\sum_{j=1}^N\int_0^t d\tau \frac{e^{i\omega_0\tau+i\ell\phi_j(t-\tau)}}{|\mathbf{r}_j(t-\tau)-\mathbf{r}|}
\int_{-\infty}^\infty dk\sin(k|\mathbf{r}_j(t-\tau)-\mathbf{r}|)
e^{-ick\tau}.
 \end{eqnarray}
Using the identity
\begin{eqnarray}
\int_{-\infty}^\infty dk \sin(kR_j)e^{-ick\tau}&=&\frac{\pi}{ic}\left[\delta(\tau-R_j/c)-\delta(\tau+R_j/c)\right],\nonumber
\end{eqnarray}
we then obtain
\begin{eqnarray}
 E_s(\mathbf{r},t)&=&\frac{dk_0^3}{4\pi\epsilon_0}\frac{\Omega_0}{2\Delta_0}R(\rho)e^{-i\omega_0t}\sum_{j=1}^N
 \frac{e^{ik_0R_j+i\ell\phi_j}}{ik_0R_j},
 \end{eqnarray}
where  $R_j=|\mathbf{r}_j-\mathbf{r}|$ and $\phi_j$ are evaluated at the retarded time $t-R_j/c$.
Assuming $\mathbf{r}\gg\mathbf{r}_j$, we can write $R_j\approx r-\hat\mathbf{n}\cdot\mathbf{r}_j$ with $\hat \mathbf{n}=\mathbf{r}/r$, so that
\begin{eqnarray}
 E_s(\mathbf{k},t)&\approx &\frac{dk_0^2}{4\pi\epsilon_0}\frac{\Omega_0}{2\Delta_0}R(\rho)\frac{e^{i(k_0r-\omega_0t)}}{ir}\sum_{j=1}^N e^{-i\mathbf{k}\cdot\mathbf{r}_j+i\ell\phi_j},
 \end{eqnarray}
where $\mathbf{k}=k_0\hat\mathbf{n}$.
At a short detection distance, we can neglect the retarded time $r/c$ and assume $\mathbf{r}_j(t-R_j/c)\approx \mathbf{r}_j(t)$ and $\phi_j(t-R_j/c)\approx \phi_j(t)$.
Thus, the angular distribution of the scattered intensity detected at a distance $r$, and in the far-field limit, is
\begin{eqnarray}
 I_s(\mathbf{k})&=&I_1N^2 |M(\mathbf{k},t)|^2,
 \end{eqnarray}
where $I_1=(\hbar\omega_0\Gamma/8\pi r^2)(\Omega_0/2\Delta_0)^2R^2(\rho)$ is the single-atom Rayleigh scattering intensity, and
\begin{eqnarray}
M(\mathbf{k},t)&=&\frac{1}{N}\sum_{j=1}^N e^{-i\mathbf{k}\cdot\mathbf{r}_j(t)+i\ell\phi_j(t)}
 \end{eqnarray}
is the dimensionless electric field emitted in direction of $\mathbf{k}$. In polar coordinates, 
$\mathbf{k}=k_0(\sin\theta\cos\phi,\sin\theta\sin\phi,\cos\theta)$ and
\begin{eqnarray}
M(\theta,\phi)&=&\frac{1}{N}\sum_{j=1}^Ne^{-ik_0\rho\sin\theta\cos(\phi-\phi_j)+i\ell\phi_j}\label{bunching:phi}.
 \end{eqnarray}
For the matter wave distribution considered in our work, it rewrites as
\begin{eqnarray}
M(\theta,\phi)&=&\frac{1}{2\pi}\int_{0}^{2\pi}e^{-ik_0\rho\sin\theta\cos(\phi-\phi')+i\ell\phi'}|\Psi(\phi')|^2d\phi'.\label{eq:5}
 \end{eqnarray}

\section{Derivation of Eq.(8)}

Let us consider the atoms initially uniformly distributed in phase on the ring, $\Psi(\phi,\tau)=(1/\sqrt{2\pi})c_0^{(\mathrm{eq})}(\tau)$.
Then, the equilibrium solution of Eq.~(4) of the main text is $c_n^{(\mathrm{eq})}(\tau)=\delta_{n0}c_0^{(\mathrm{eq})}(\tau)$, where
\begin{equation}
c_0^{(\mathrm{eq})}(\tau)=e^{-i(\gamma/2)V_0\tau}.
\end{equation}
The stability of this equilibrium point can be determined by a linear stability analysis, perturbing the equilibrium by
\begin{equation}
c_n(\tau)=\left[\delta_{n0}+\delta c_n(\tau)\right]e^{-i(\gamma/2)V_0\tau}.
\end{equation}
At the first order in the perturbation $\delta c_n$, $\sum_n c^*_{n-k}c_n \approx \delta c_k+\delta c^*_{-k}$
and from Eq.~(4) in the main text we obtain
\begin{eqnarray}
\frac{d}{d\tau}\delta c_m&=&-im^2\delta c_m-i\frac{\gamma}{2}V_m(\delta c_m+\delta c^*_{-m}),\\
\frac{d}{d\tau}\delta c_{-m}^*&=&im^2\delta c_{-m}^*+i\frac{\gamma}{2}V_{m}(\delta c_m+\delta c^*_{-m}),
\end{eqnarray}
where we used the property $V_m^*=V_{-m}$.
For each Fourier component $V_m$ which is different from zero, the transition occurs from $m=0$ toward the states $m$ and $-m$.
Defining $\Phi_m=\delta c_m+\delta c^*_{-m}$ and $\Pi_m=-i(\delta c_m-\delta c^*_{-m})$, we obtain
\begin{eqnarray}
\frac{d}{d\tau}\Phi_m&=& m^2\Pi_m,\\
\frac{d}{d\tau}\Pi_m&=& -\left(m^2+\gamma V_m\right)\Phi_m.
\end{eqnarray}
This leads to
\begin{eqnarray}
\frac{d^2}{d\tau^2}\Phi_m&=& \lambda_m^2\Phi_m,
\end{eqnarray}
where $\lambda_m^2=-m^2(m^2+\gamma V_m)$ and
\begin{equation}
\lambda_m=\pm i m\sqrt{m^2+\gamma V_m}.
\end{equation}

\bibliography{Bibliography}